\documentclass[aps,prb,twocolumn,groupedaddress]{revtex4-1}
\usepackage{amsmath}
\usepackage{graphicx}
\usepackage{epstopdf,amssymb}
\usepackage{dcolumn}% Align table columns on decimal point
\usepackage{bm}
\usepackage{color}
\def\be{\begin{equation}}
\def\ee{\end{equation}}
\def\bea{\begin{eqnarray}}
\def\eea{\end{eqnarray}}

\begin{document}
\title{Phonon assisted resonant tunnelling and its phonons control}
\author{F.\,V.\,Kusmartsev}
\email{F.Kusmartsev@lboro.ac.uk}
\affiliation{Department of Physics,
Loughborough University, LE11 3TU Loughborough, United Kingdom}
\author{V.\,D.\,Krevchik}
\affiliation{Department of Physics, Penza State University, 440026
Penza, Russia}
\author{M.\,B.\,Semenov}
 \email{misha29.02@gmail.com}
\affiliation{Department of Physics, Penza State University, 440026
Penza, Russia}
\author{D.\,O.\,Filatov}
\affiliation{Lobachevskii University of Nizhniy Novgorod, 603950
Nizhniy Novgorod, Russia}
\author{A.\,V.\,Shorokhov}
\affiliation{Mordovia State University, 430005 Saransk, Russia}
\author{A.\,A.\,Bukharaev}
\affiliation{Zavoisky Institute for Physics and Technology, Kazan
Scientific Center, Russian Academy of Science,  420029 Kazan, Russia
and Kazan Federal University, 420008 Kazan, Russia}
\author{Yuri~Dakhnovsky}
\affiliation{Department of Physics \& Astronomy, University of
Wyoming, WY 82071 Laramie, USA}
\author{A.\,V.\,Nikolaev}
\affiliation{Skobeltsyn Institute of Nuclear Physics, Lomonosov
Moscow State University, 119991 Moscow, Russia and Moscow Institute
of Physics and Technology, 141700 Dolgoprudny, Russia}
\author{N.\,A.\,Pyataev}
\affiliation{Mordovia State University, 430005 Saransk, Russia}
\author{R.\,V.\,Zaytsev}
\affiliation{Department of Physics, Penza State University, 440026
Penza, Russia}
\author{P.\,V.\,Krevchik}
\affiliation{Department of Physics, Penza State University, 440026
Penza, Russia}
\author{I.\,A.\,Egorov}
\affiliation{Department of Physics, Penza State University, 440026
Penza, Russia}
\author{K.\,Yamamoto}
\affiliation{Research Institute, International Medical Center,
2-25-22-304 Kohinata Bunkyo-ku, Tokyo, Japan}
\author{A.\,K.\,Aringazin}
\affiliation{Institute for Basic Research, Eurasian National
University, 010008 Astana, Kazakhstan}

%\date{\today}

\begin{abstract}
We observe a series of sharp resonant features in the tunnelling
differential conductance of InAs quantum dots. We found that
dissipative quantum tunnelling has a strong influence on the
operation of nano-devices. Because of such tunnelling the
current-voltage characteristics of tunnel contact created between
atomic force microscope tip and a surface of InAs/GaAs quantum dots
display many interesting peaks. We found that the number, position,
and heights of these peaks are associated with the phonon modes
involved. To describe the found effect we use a quasi-classical
approximation. There the tunnelling current is related to a creation
of a dilute instanton-anti-instanton gas. Our experimental data are
well described with exactly solvable model where one charged
particle is weakly interacting with two promoting phonon modes
associated with external medium. We conclude that the
characteristics of the tunnel nanoelectronic devices can thus be
controlled by a proper choice of phonons existing in materials,
which are involved.
\end{abstract}
 \maketitle

%Key words: quantum tunneling with dissipation, quantum dots, tunnel current-voltage characteristics, conducting atomic force microscopy.

%\PACS{03.65.Xp, 31.15.xg, 73.40.Gk, 82.20.Xr}

\section{Introduction} \label{Sec1}
One of the practical problems in semiconductor tunnel
nanoelectronics is to extend control over parameters of quantum
tunnel effect of electrons [\onlinecite{2}-\onlinecite{19}].
Physical and chemical approaches to electron transfer processes at
nanoscales reveal some common tools. Namely, analytical models of
multidimensional dissipative simultaneous quantum tunneling of one
or two charged particles, electrons or protons, in low-dimensional
systems used in studying some chemical reactions at low temperatures
[\onlinecite{9}-\onlinecite{Dakhnovskii88}], can be used to study
physical properties of quantum dots (QDs) [\onlinecite{10}],
[\onlinecite{20}-\onlinecite{22}].

Quasiclassical, dilute instanton-antiinstanton gas approach to the dissipative quantum tunneling of particles interacting
 with heat bath is known to be powerful technique in obtaining analytical
 results, starting from a classical action of the system [\onlinecite{2}-\onlinecite{4}]. In particular,
  probability rate of quantum tunneling of two mutually interacting charged
  particles moving in medium, which occurs in synchronous or asynchronous modes,
   was studied in our previous work [\onlinecite{7}].

The purpose of the present work is to identify experimentally
observed dissipative tunneling effects predicted by the theory
developed in the pioneering works of A.J. Leggett, A.I. Larkin,
Yu.N. Ovchinnikov et al. In this paper, we show results of our
recent experimental study of the effect of wide-band matrix on
semiconductor InAs/GaAs(001) QDs which changes some macroscopic
properties of the system, and can be identified by the tunnel
current-voltage (\textit{I---V}) relationship of the tunnel device.
Parameter associated to this effect can be treated as one more
controllable parameter of nanostructures, in addition to energy
levels of QD which are controlled by its size parameter.  Note that
the role of wide-band matrix in taking control over mesoscopic
systems has been stressed in [\onlinecite{1}]. Also, we develop and
apply 1D dissipative tunneling model accounting for influence of
\textit{two }promoting phonon modes coming from the wide-band
matrix, to study quantum tunneling in the structure of single QDs by
Conducting Atomic Force Microscopy (CAFM). We make a comparison of
the obtained theoretical tunneling probability rate in an
oscillatory regime with the experimental \textit{I---V} relationship
of the contact between Atomic Force Microscopy (AFM) probe and
InAs/GaAs(001) QD surface.

\section{1D dissipative tunneling probability rate with account for
two local phonon modes in the wide-band matrix}\label{Sec2}

 We start
by short description of the dissipative tunneling approach, which
will be used in formulating our specific model. Let $p_1$ is the
momentum of the tunneling particle, $y_1$ is the coordinate of the
the tunneling particle, $v_1$ is the two-well potential. Then the
Hamiltonian
\begin{equation} \label{GrindEQ__1_}
\hat H=\frac{p_{1} ^{2} }{2} +v_{1} (y_{1} )+y_{1} \sum _{\alpha =2}^{N}C_{\alpha } y_{\alpha } +\frac{1}{2}
 \sum _{\alpha =2}^{N}\left(p_{\alpha } ^{2} +\omega _{\alpha } ^{2} y_{\alpha } ^{2} \right)
\end{equation}
describes particle in the model asymmetric two-well oscillatory
potential $v_1$ along the tunnel coordinate $y_1$.  In
Eq.(\ref{GrindEQ__1_}) $p_{\alpha }$ are momenta of phonon modes of
the particle of mass $m=1$, $y_{\alpha }$ are coordinates of local
phonon modes,
 $ {\omega }_{\alpha }$ are frequencies of local phonon modes,  $N$ is
 the number of local modes of wide-band matrix, and $C_{\alpha }$ are coefficients of the interaction of the tunneling
 particle with local phonon modes of wide-band matrix (see [\onlinecite{13}, \onlinecite{7},\onlinecite{Dakhnovskii88}] for more details).

%%%%%%%%%%%%%%%%%%%%%%%%%%%%%%%%%%%%%%%%%%%%%%%%%%%%%%
\begin{figure}[h]
\begin{center}
\includegraphics[clip=true,width=0.7\linewidth]{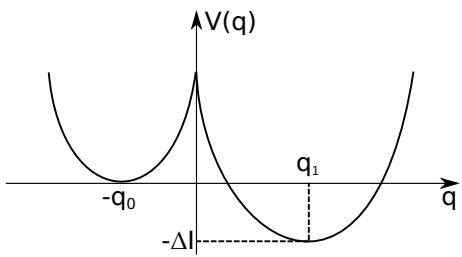}
\end{center}
\caption{ \label{Fig1} Asymmetric two-well oscillatory potential
along the tunnel coordinate $q\ $of the model.}
\end{figure}
%%%%%%%%%%%%%%%%%%%%%%%%%%%%%%%%%%%%%%%%%%%%%%%%%%%%%%

The quasiclassical action ${S\{y}_1\}$ of the system is defined by
[\onlinecite{7}]
\begin{eqnarray}
S\left\{y_{1} \right\}=\int _{-\beta /2}^{\beta /2}d\tau
\left[\frac{1}{2}\dot{y}_{1}^2 +v(y_{1} ) \right.
 \nonumber \\
 \left.+\frac{1}{2} \int _{-\beta /2}^{\beta /2}d\tau '{\kern 1pt} K\left(\tau -\tau '\right){\kern 1pt} {\kern 1pt} y_{1} (\tau )y_{1} (\tau ')
 \right].
\label{GrindEQ__2_}
\end{eqnarray}
Hereafter, dot denotes derivative over $\tau$, $\beta =2\pi
h{/k}_BT$ is the inverse temperature,
$K(\tau)=T\sum_{\omega_n}K(\omega_n)\exp(-i\omega_n\tau)$ is the
Matsubara Green's functions [\onlinecite{2},\onlinecite{3a}],
$K(\omega_n)=-\sum_\alpha C_\alpha^2/(\omega_\alpha^2+\omega_n^2)$,
$\omega_n=2\pi Tn$, and we introduced the renormalized potential
$v(y_1)$
\begin{equation} \label{GrindEQ__3_}
v(y_{1} )=v_{1} (y_{1} )-\frac{1}{2} \sum _{\alpha =2}^{N}\frac{C_{\alpha } ^{2} }{\omega _{\alpha } ^{2} } {\kern 1pt} {\kern 1pt} {\kern 1pt} y_{1} ^{2} .
\end{equation}
The form of the potential $v$ as a function of the renormalized
coordinate $q$ is shown in Fig.~\ref{Fig1}. The procedures of the
renormalization of the potential $v_1(y_1)$ and the coordinate $y_1$
are considered in [\onlinecite{Dakhnovskii88}].

 In the instanton approximation, 1D Euclidean action $S_B$ for
one charged particle in the two-well renormalized oscillator
potential and the external electric field $E$ is found as
[\onlinecite{13}, \onlinecite{14}, \onlinecite{20}] \bea S_{B}
=2\omega _{0} ^{2} \left(q_{0} +q_{1} \right)q_{0} \tau _{0}
-\frac{2\omega _{0} ^{2} \left(q_{0} +q_{1} \right)^{2} \tau _{0}
^{2} }{\beta }
\nonumber\\
-\frac{4\omega _{0} ^{4} \left(q_{0} +q_{1} \right)^{2} }{\beta } \sum _{n=1}^{\infty }\frac{\sin ^{2} \nu _{n} \tau _{0}^{} }{\nu _{n} ^{2} \left(\nu _{n} ^{2} +\omega _{0} ^{2} +\zeta _{n} \right)}  ,
\label{GrindEQ__4_}
\eea
where $e$ is charge of the particle, $q_{0} =b^{*} -{|{\kern 1pt} e{\kern 1pt} |E}/{\omega _{0}^{2} } $ and $q_{1} =b^{*} +{|{\kern 1pt} e{\kern 1pt} |E}/{\omega _{0}^{2} } $ are parameters of the renormalized two-well potential in the external electric field $E$, $\pm b^*$ are positions of the minima of the potential in zero-field case, ${\tau }_0$ is the instanton center, ${\omega }_0$ is the oscillator potential frequency, ${ \nu }_n=2\pi n/\beta $ is Matsubara frequency, $n=1,2,3,\dots $,  and $ {\zeta }_n$ is the Fourier  component of the viscous core of the corresponding quasiclassical Euler-Lagrange equation of motion,
\begin{equation} \label{GrindEQ__5_}
\zeta _{n} =\nu _{n} ^{2} \sum _{\alpha =2}^{N}\frac{C_{\alpha } ^{2} }{\omega _{\alpha } ^{2} \left(\omega _{\alpha } ^{2} +\nu _{n} ^{2} \right)}.
\end{equation}

\begin{figure}[h]
\begin{center}
\includegraphics[clip=true,width=0.7\linewidth]{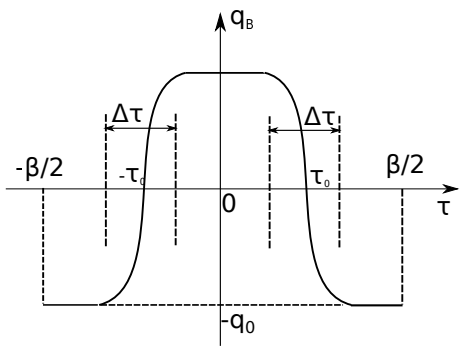}
\end{center}
\caption{\label{Fig2} The instanton $q_B\left(\tau \right)$, for
${\tau } >0$. ${\tau }_0$ and $\Delta \tau \ $ are the so called
center and width of the instanton, respectively. }
\end{figure}

The probability of quantum tunneling of the particle through the
barrier shown in Fig.\ref{Fig1} per unit time $\Gamma =B\exp
\left(-S_{B} \right)$ contains pre-exponential factor $B$, the most
contribution to which is made by the particle  trajectories that are
very close to the instanton. Expanding the action up to quadratic
term in ${q-q}_B$, where $q_B$ is the extremal instanton trajectory,
and integrating over the functional space, we get
\begin{equation} \label{GrindEQ__7_}
B=\left[\frac{S_{0} }{2\pi } \, \cdot \, \frac{\det \left(\frac{\delta ^{2} S}{\delta q^{2} } \right)_{q=-q_{0} } }{\det'\left(\frac{\delta ^{2} S}{\delta q^{2} } \right)_{q=q_{B} (\tau )} } \right]^{1/2} ,
\end{equation}
where
\begin{equation} \label{GrindEQ__8_}
S_{0} =\int _{-\beta /2}^{\beta /2}\dot{q}_{B} ^{2} (\tau )\, d\tau  .
\end{equation}
Here, ${\det}'$ denotes determinant with zero eigenvalues corresponding to zero modes of the instanton omitted, and we used the approximation of dilute instanton-antiinstanton gas, i.e., the probability rate is much smaller than the inverse width of instanton,
\begin{equation} \label{GrindEQ__9_}
\Gamma \ll\left(\Delta \tau \right)^{-1} ,
\end{equation}
We assume that the major contribution to the action $S\{q\}$ is made by the instanton, i.e., by the trajectory $q_B(\tau )$, which minimizes the action (\ref{GrindEQ__4_}) and obeys the following Euler-Lagrange equation:
\begin{equation} \label{GrindEQ__10_}
-\ddot{q}_{B} (\tau )+\frac{\partial {\kern 1pt} v(q_{B} )}{\partial {\kern 1pt} q_{B} } +\int _{-\beta /2}^{\beta /2}d\tau '{\kern 1pt} {\kern 1pt} K\left(\tau -\tau '\right){\kern 1pt} {\kern 1pt} q_{B} \left(\tau '\right) =0.
\end{equation}
Here, the trajectory $q_B(\tau )$ is found within the class of
periodic functions $q_{B} (\tau )=q_{B} (\tau +\beta )$. The form of
the solution of Eq.(\ref{GrindEQ__10_}) obeying this condition is
shown in Fig.~\ref{Fig2}.

%% FIGURE 4
%%%%%%%%%%%%%%%%%%%%%%%%%%%%%%%%%%%%%%%%%%%%%%%%%%%%%%%
%%%%%%%%%%%%%%%%%%%%%%%%%%%%%%%%%%%%%%%%%%%%%%%%%%%%%%%

With the action (\ref{GrindEQ__4_}), we obtain $B$ from Eq. (\ref{GrindEQ__7_}) as follows:
\begin{equation} \label{GrindEQ__12_}
B=\frac{2\omega _{0} ^{2} \left(q_{0} +q_{1} \right)^{2} }{\left(2\pi \beta \right)^{1/2} }\!\!\!\! \sum _{n=-\infty }^{\infty }\!\!\frac{\sin ^{2} \nu _{n} \tau _{0} }{\lambda _{0n} } \left(\sum _{n=-\infty }^{\infty }\frac{\cos 2\nu _{n} \tau _{0} }{\lambda _{0n} }  \right)^{\!\!-1/2}.
\end{equation}
Now, we specify our model by assuming that the particle interacts
weakly with two local phonon modes, i.e., $\omega _{L1} =\omega _{2}
$ and $\omega _{L2} =\omega _{3} $.

 %in contrast  to our previous
%study \cite{12} when only one local phonon mode was taken into
%account.

We also make natural assumption that the interactions are weak, i.e., ${C_{\alpha } }/{\omega _{0} ^{2} } \ll 1$ and ${C_{\alpha } }/{\omega _{L} ^{2} } \ll 1$. In this case, we get
\be
\zeta _{n} =\nu _{n}^{2} \frac{C_{2}^{2} }{\omega _{2}^{2} (\omega _{2}^{2} +\nu _{n}^{2} )} +\nu _{n}^{2} \frac{C_{3}^{2} }{\omega _{3}^{2} (\omega _{3}^{2} +\nu _{n}^{2} )}.
%\quad
%%\sin^{2} \nu _{n} \tau _{0} =\frac{1}{2} (1-\cos\, 2\nu _{n} \tau _{0} ).
 \label{GrindEQ__13_}
\ee The tedious calculations  provides an analitic formulae for the
optimal action $\tilde S_B$ and pre-exponetial factor $B$. Due its
large  expression it is given in [\onlinecite{Suppl}].
For a completeness, we should mention that there is a second
solution which corresponds to the case of non-oscillating character
of the probability rate. We shortly present the results for this
case in [\onlinecite{Suppl}].

\section{Experimental setup}\label{Sec3}

\begin{figure}[h]
\begin{center}
\includegraphics[clip=true,width=0.8\linewidth]{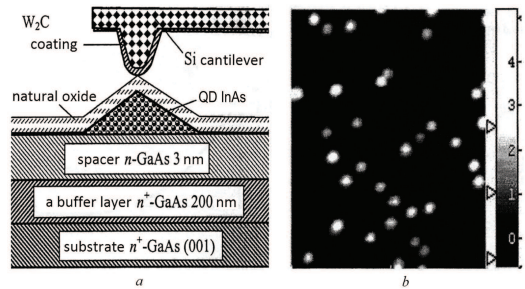}%
\end{center}
\caption{\label{Fig3} (a)  Schematic representation of the
experimental setup for measuring electric current
 between AFM probe and the conductive substrate via surface of InAs/GaAs(001) QD by CAFM; (b)
 AFM image of the surface of InAs/GaAs(001) QD structure; 750 by 700 nm size, range of heights is 5.9 nm.
}
\end{figure}
%%%%%%%%%%%%%%%%%%%%%%%%%%%%%%%%%%%%%%%%%%%%%%%%%%%%%%

The experimental setup is shown in Fig.~\ref{Fig3}. Samples for the
study were prepared on the n$^+$-GaAs(001) substrate doped by Sn
using Metalorganic Hydridechemical Vapor Deposition (MOCVD) at
atmospheric pressure.

 The n$^+$-GaAs buffer layer of 200 nm thickness doped by Si, with the
donor concentration $N_D \simeq 10^{18}$ cm$^{-3}$ has been grown at
$T=650$ C. The \textit{n}-GaAs spacer layer, with $N_{D} \simeq
10^{15}$ cm$^{-3}$, of 3 nm thickness has been grown on the buffer
layer. This layer forms triangular potential barrier between the QD
and n$^+$-GaAs buffer layer [\onlinecite{25}]. InAs QD has been
formed by Stransky-Krastanov mechanism at $T=530$ C. Nominal
thickness of deposited InAs layer is about 1.5 nm.

The samples have been prepared in Physical-Technical Research Institute,
Nizhniy Novgorod University, Russia, and then used to study spatial and
 energy distribution of the local density of states in InAs QD by the method
 of tunnel AFM, in Kazan Physical-Technical Research Institute, Kazan, Russia.

Note that in [\onlinecite{11a}] local density of states in
InAs/GaAs(001) QDs was measured by CAFM. In [\onlinecite{11c}],
Scanning Tunneling Microscopy (STM) has been used in Ultra-High
Vacuum (UHV) to measure local density of states in quantum wells
GaSb/InAs. In [\onlinecite{11b}], \textit{combined} UHV STM/AFM has
been implemented for the first time for the tunneling spectroscopy
of the size-quantized states in the InAs/ GaAs (001) surface QDs.
Tunneling spectra and current images, which reflect the energy and
spatial distribution of the local density of the ground and excited
states in the QDs have been obtained. Tunneling AFM technique and
results of \textit{ex situ} investigations of local density of
states of quantum confined states in self-assembled semiconductor
InAs/GaAs(001) QDs and InGaAs/GaAs InAs/GaAs(001) quantum rings,
grown by Atmospheric Pressure Metal Organic Vapour Phase Epitaxy
(AP-MOVPE), and GeSi/Si InAs/GaAs(001) nanoislands covered by native
oxide have been presented in [\onlinecite{23}]. There, samples with
surface nanostructures were scanned across by conductive Si AFM
probe covered by conductive coating (Pt, W$_2$C, or diamond-like
film) in the contact mode. Main advantage of the tunneling AFM as
compared to UHV STM is that the former allows \textit{ex situ}
investigation of the surface semiconductor nanostructures, which are
naturally oxidized in ambient air when one takes samples  from
growth setup to AFM setup.

 Our experiment was done at room
temperatures, under very high vacuum in the camera, by scanning
probe microscope \textit{Omicron UHV AFM/STM VT} which is a part of
very high vacuum set up \textit{Omicron MultiProbe P}. Basic
pressure in the camera was about 10$^{-10}$ Torr. Surface of the
sample experienced oxidation in ambient air during the time of
transportation from the growth set up to the vacuum camera. It was
scanned by p$^+$-Si probe with W$_{2}$C covering, in contact mode
(see Fig.~\ref{Fig3}), with potential difference $V_g$ between
n$^+$-GaAs substrate and AFM probe.

 In the experiment, we have obtained spatial distributions of electric current $I_t$ between AFM probe and the sample as a function
 of AFM probe coordinates $(x,\ y)$ in the plane of sample's surface,
  at various constant $V_g$. We refer to these distributions as current images.
   They reflect spatial distribution of local density of states in the plane: wave function mappings
   [\onlinecite{11c}] with a sum over energies below Fermi level of the probe. We do not present resulting current images here,
   and mention only that for the potential difference range $V_g = 2.6 \dots 3.1$ V across it, these reveal two maxima corresponding
   to $p$-symmetry of the excited states of QDs, while at lower voltages $V_g < 2.6$ V the current images have round form corresponding to the $s$-symmetry of the ground state
   [\onlinecite{24b}].

  \textit{I---V} relationship of the contact between the AFM probe and QD have been obtained by measuring
  current images at different fixed $V_g$. More details on the used method of growth and tunnel spectroscopy of QD one can find, e.g., in [\onlinecite{24}].

Fig.\ref{Fig3}b represents AFM image of the surface of the studied
sample.
 surface QD has 5 to 6 nm thickness. It should be noted that lateral dimensions of QD shown in Fig.\ref{Fig3}b are significantly larger than that expected for QD, 10 to 12 nm,
 which has form of tetrahedral pyramid with planes (101). This is
 explained by the effect of convolution related to the curvature of
  the AFM probe characterized by the radius $R_p \simeq 35$ nm.

\section{Comparison with the experiment}\label{Sec4}

We turn to a comparison of the obtained theoretical results for the 1D
quantum tunneling probability rate of the charged particle weakly interacting
 with the two local phonon modes of medium with the experimental \textit{I---V} relationship of InAs/GaAs(001) QD measured by using CAFM.

\begin{figure}[h]
\begin{center}
\includegraphics[clip=true,width=0.8\linewidth]{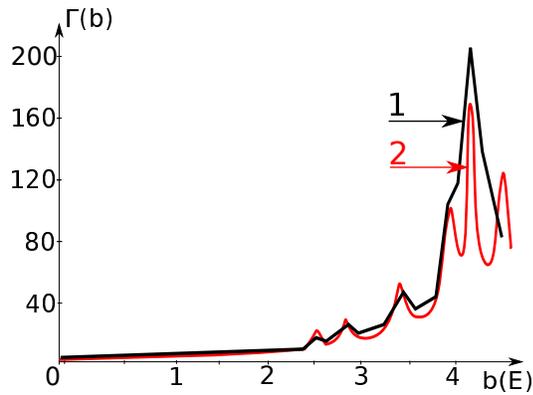}
\end{center}
\caption{ \label{Fig4} Comparison of the theoretical result
(oscillatory case) for the 1D quantum tunneling probability rate
$\Gamma$ of the charged particle weakly interacting with the two
local phonon modes of medium (curve 1) as a function of the symmetry
parameter $b$ lineraly depending on the external electric field
intensity $E$ with the experimental \textit{I---V} relationship for
the probe-to-sample contact for InAs/GaAs(001) QD at 300 K
(represented as curve 2) obtained in the conductive atomic force
microscopy experiment. Shown numerical values on the axes are in the
units of applied voltage varied in the range $V_g =0.0 \dots 4.0 $~V
with the step 100~mV and the accuracy 1~mV, and the current $I$, in
pA, measured with the accuracy 10 pA. }
\end{figure}

In Fig.~\ref{Fig4}, we plot the experimental \textit{I---V}
relationship and the obtained result for tunneling probability
 rate $\Gamma$, which we treat as proportional to the current $I$, as a function of the parameter $b = q_0$ linearly depending
  on the external electric field intensity $E$, which in turn linearly depends on the potential difference $V$ across it.

Fitting parameters are the frequencies of local phonon modes and the coupling constants of the particle to these modes.
Values of other parameters of the model, namely, the two-well potential parameters, temperature, and external electric field intensity were set due to the experiment.

The observed peaks are not due to Coulomb blockage because these are not equidistant. Sometimes these peaks are interpreted as the result of a resonance phenomenon in tunneling.

While the position of the highest theoretical peak meets the experimental one, its height underestimates
the latter by about 20\%. Also, we note that the highest experimental peak reveals signs of two additional
 peaks matching two theoretical ones close to the highest peak. This can be explained by
 an approximate character of the used asymmetric two-well oscillatory potential.

The case of non-oscillating character of the probability rate
obviously does not correspond to the experimental data.
Nevertheless, we present shortly the results in Supplemental
Material [\onlinecite{Suppl}]. This damping solution is valuable as
it tells us that whether the rate is of oscillating or
non-oscillating character depends on the temperature, external
electric field intensity $E$, and type of wide-band matrix, which
carries studied QDs.

%% FIGURE 5
%%%%%%%%%%%%%%%%%%%%%%%%%%%%%%%%%%%%%%%%%%%%%%%%%%%%%%%

\section{Conclusions}\label{Sec5}

Our measurements reveal a rich spectrum of inelastic phonon-assisted
electron tunneling processes in InAs/GaAs quantum dots. The number,
position, and heights of the resonant peaks observed is well
described with developed instanton approach extended to include
effects of dissipative quantum tunneling. The good agreement with
experiments have been achieved using a model in which an inelastic
tunneling transition becomes allowed. There in the model of the
wide-band matrix of InAs/GaAs quantum dots two phonon modes have
been taken into account. The comparison between the experiments and
developed theory allows us to probe electron-phonon interactions in
this system and identify spectroscopically the energies and nature
of the phonons emitted during the tunnelling.

Practical implication of the obtained result is that the
current-voltage characteristics of the semiconductor tunnel
nanoelectronic devices can thus be controlled to a certain extent by
modulations of the wide-band matrix parameters. For example, the
quantum dots can serve as very sensitive detectors of properties of
bulk materials used in precision nanoelectronics.

 It is necessary to
note that the theory for dissipative tunneling with the influence of
two local phonon modes in the external electric field developed in
this paper can be used in problems of resonant impurity states in
quantum molecules in the case when the lifetime of the impurity
electron is mainly determined by electron tunneling decay.

The problem of the influence of the electric field on the radiative
recombination spectra of electrons and holes in a quantum dot is
also of interest [\onlinecite{Zegrya2006}-\onlinecite{Zegrya2009}].
In this case the electron tunneling between a quantum dot and
biological object can play an important role.

%%%%%%%%%%%%%%%%%%%%%%%%%%%%%%%%%%%%%%%%%%%%%%%%%%%%%%%%%

%% FIGURE 6
%%%%%%%%%%%%%%%%%%%%%%%%%%%%%%%%%%%%%%%%%%%%%%%%%%%%%%%
%\begin{figure}[h]
%\begin{center}
%\includegraphics[clip=true,width=0.8\linewidth]{fig6.eps}
%\end{center}
%\caption{ \label{Fig6} Fig 6. Theoretical result for the 1D quantum
%tunneling probability rate $\Gamma$ of the charged particle weakly
%interacting with the two local phonon modes of medium as a function
%of the symmetry parameter $b$ lineraly depending on the external
%electric field intensity $E$ (non-oscillatory case).  The same units
%are used as in Fig.~5.}
%\end{figure}
%%%%%%%%%%%%%%%%%%%%%%%%%%%%%%%%%%%%%%%%%%%%%%%%%%%%%%%

 A.K.A., M.B.S., and V.D.K. are grateful to Committee of Science of the Ministry of Education and Science of
 Kazakhstan for a partial financial support under the grant.
The part of this work was done in the framework of the state
contract by Ministry of Education and Science of the Russian
Federation. The authors are also grateful to A. J. Leggett for his
attention to this work; to B.N. Zvonkov, Research Institute for
Physics and Technology, N.I. Lovachevskii University of Nizhny
Novgorod, for QD samples used in CAFM experiment; to P.A. Borodin
for assistance in carrying out CAFM experiment, and to I.E.
Bulyzhenkov and Yu.N. Ovchinnikov for useful discussions.

\end{document}